\documentclass[conference]{IEEEtran}
\IEEEoverridecommandlockouts
\usepackage{cite}
\usepackage{amsmath,amssymb,amsfonts}
\usepackage{algorithmic}
\usepackage{graphicx}
\usepackage{textcomp}
\usepackage{xcolor}
\usepackage{verbatim}
\usepackage{mathtools, nccmath}
\usepackage{multirow}
\usepackage{array}

\ifCLASSOPTIONcompsoc
    \usepackage[caption=false, font=normalsize, labelfont=sf, textfont=sf]{subfig}
\else
\usepackage[caption=false, font=footnotesize]{subfig}

\def\BibTeX{{\rm B\kern-.05em{\sc i\kern-.025em b}\kern-.08em
    T\kern-.1667em\lower.7ex\hbox{E}\kern-.125emX}}
\begin{document}
\setlength{\abovedisplayskip}{3pt}
\setlength{\belowdisplayskip}{3pt}

This work has been submitted to the IEEE for possible publication. Copyright may be transferred without notice, after which this version may no longer be accessible.
\newpage
\title{An overview of physical layer design for Ultra-Reliable Low-Latency Communications in 3GPP Release 15 and Release 16\\
}

\author{\IEEEauthorblockN{Trung-Kien Le$^{\dagger}$, Umer Salim$^{\star}$, Florian Kaltenberger$^{\dagger}$}
\IEEEauthorblockA{$^{\dagger}$EURECOM, Sophia-Antipolis, France \\
$^{\star}$TCL Mobile, Sophia-Antipolis, France \\
Trung-Kien.Le@eurecom.fr}
}

\maketitle

\begin{abstract}

For the fifth generation (5G) wireless network technology, the most important design goals were improved metrics of reliability and latency, in addition to network resilience and flexibility so as to best adapt the network operation for new applications such as augmented virtual reality, industrial automation and autonomous vehicles. That led to design efforts conducted under the subject of ultra-reliable low-latency communication (URLLC). The technologies integrated in Release 15 of the 3rd Generation Partnership Project (3GPP) finalized in December 2017 provided the foundation of physical layer design for URLLC. New features such as numerologies, new transmission schemes with sub-slot interval, configured grant resources, etc., were standardized to improve mainly the latency aspects.

5G evolution in Release 16, PHY version finalized in December 2019, allows achieving improved metrics for latency and reliability by standardizing a number of new features. This article provides a detailed overview of the URLLC features from 5G Release 15 and Release 16, describing how these features permit URLLC operation in 5G networks.

\end{abstract}


\section{Introduction} \label{I}


To satisfy the requirements of emerging applications such as intelligent transportation, augmented virtual reality, industrial automation, etc, 3GPP defined three new service categories: enhanced mobile broadband (eMBB), massive machine-type communication (mMTC) and ultra-reliable low-latency communication (URLLC). The physical design of URLLC is the most challenging one because two conflicting factors of reliability and latency have to be coped with at the same time.

In \cite{ad31}, the terms reliability and latency are defined as follows: ``Reliability can be evaluated by the success probability of transmitting X bytes within a certain delay, which is the time it takes to deliver a small data packet from the radio protocol layer 2/3 SDU ingress point to the radio protocol layer 2/3 SDU egress point of the radio interface, at a certain channel quality (e.g., coverage-edge)''. It is measured by block error rate (BLER) using Monte-Carlo simulations. In \cite{ad32}, 3GPP defines targets for the URLLC scenario: ``A general URLLC reliability requirement for one transmission of a packet is 10\textsuperscript{-5} for 32 bytes with a user plane latency of 1 ms''. The next release of 3GPP will have higher requirements of URLLC: ``Higher reliability (up to 10\textsuperscript{-6}), short latency in the order of 0.5 to 1 ms, depending on the use cases (factory automation, transport industry and electrical power distribution)''\cite{ad33}. 

3GPP Release 15 built a foundation for URLLC design to achieve these stringent requirements by introducing higher subcarrier spacings and thus shorter OFDM symbol lengths (this is also called numerology), sub-slot transmission time intervals, or configured grant resources (a summary of these features is presented in Section \ref{IA}). Release 16 continues to develop the physical layer design for URLLC further to satisfy more stringent requirements in some scenarios as specified in \cite{ad33}. An overview of the challenges and the techniques standardized in Release 16 are described in Section \ref{II}. Some conclusions are drawn in Section \ref{III}.

\section{Techniques standardized in 3GPP Release 15}\label{IA}

To achieve the strict requirements of reliability and latency, 3GPP standardized several techniques in physical layer design in Release 15 finalized in December 2018.

\subsection{5G flexible numerlogy}

New numerology is specified when the values of subcarrier spacing (SCS) are 15 kHz, 30 kHz, 60 kHz, 120 kHz and 240 kHz \cite{ad26} instead of 15 kHz in LTE to reduce the time length of OFDM symbols.

Besides slot-based transmission (physical downlink shared channel (PDSCH)/physical uplink shared channel (PUSCH) mapping Type A), mini-slot based transmission (PDSCH/PUSCH mapping Type B) is also used where a packet is scheduled in the transmission occasion of 2, 4 or 7 OFDM symbols instead of an entire slot so it is transmitted immediately without waiting for slot boundary \cite{ad26}. 

\subsection{Downlink (DL) pre-emption}

In DL, the base station (gNB) can pre-empt an eMBB transmission to allocate those punctured resources to an URLLC transmission so that the URLLC packet is transmitted as soon as posible with eMBB and URLLC multiplexing instead of waiting until the end of the ongoing eMBB transmission. The gNB transmits an pre-emption indication (PI) to the eMBB UE so as to inform that the resources indicated contain data of URLLC transmission rather than its own eMBB transmission \cite{ad30}. Thus, the eMBB user (UE) does not take into account the resources punctured when decoding data.  

\subsection{Uplink (UL) configured grant (CG) transmission}

In UL, CG resources are configured to the UEs by the gNB. The UE uses these CG resources to transmit data on PUSCH directly to the gNB without transmitting scheduling request (SR) and receiving UL grant as dynamic grant (DG) transmission \cite{ad27}. There are two types of CG PUSCH transmission. In Type 1 CG PUSCH transmission, a radio resource control (RRC) signalling configures the time domain resource allocation including periodicity of CG resources, offset, start symbol and length of PUSCH as well as the number of repetitions. In Type 2 CG PUSCH transmission, only periodicity and the number of repetitions are configured by RRC signalling. The other time domain parameters are configured through an activation downlink control information (DCI). CG resources might be shared among several UEs based on contention-based access.

In UL CG transmission, the UE is configured by parameters from higher layer to transmit automatically a number of repetitions without waiting feedback from the gNB \cite{ad27}. 

Multiple CG configurations are supported for different services/traffic types and/or for enhancing reliability and reducing latency \cite{ad13}.

\section{Enhancements in physical layer design for URLLC in 3GPP Release 16}\label{II}

\subsection{Physical downlink control channel (PDCCH) enhancements}\label{IIA}

\subsubsection{PDCCH monitoring capability enhancements}\label{IIA1}

PDCCH is a physical channel that carries DCI. A PDCCH consists of 1, 2, 4, 8 or 16 control channel elements (CCE). A CCE consist of 6 resource element groups (REGs). A REG equals one resource block (RB) during one Orthogonal Frequency Division Multiplexing (OFDM) symbol that contains 12 resource elements (REs). The number of CCEs that a PDCCH has is defined as the aggregation level (AL).

DCI is transmitted from the gNB to the UE on PDCCH. The UE does not know the exact location of PDCCH so it carries out blind decoding in a search space inside the control resource sets (CORESETs). A UE is configured with a number of CORESETs to monitor PDCCH. Each of the possible location of PDCCH in the search space is called PDCCH candidate. PDCCH candidates can have overlapped CCEs. For example, a PDCCH candidate AL 8 has two CCEs overlapping with a PDCCH candidate AL 2.

Table~\ref{tab1} shows the monitoring capability of a UE defined per slot in terms of SCS as standardized in Rel-15 \cite{ad22}.

\begin{table}[htbp]
\caption{UE monitoring capability in a slot in Release 15}
\begin{center}
\begin{tabular}{|p{12em}|p{2.5em}|p{2.5em}|p{2.5em}|p{2.7em}|}
 \hline
 \textbf{\textit{SCS}} & 15kHz &30kHz &60kHz&120kHz \\ 
 \hline
 \textbf{\textit{Number of monitored PDCCH candidates}} & 44 &36 &22&20 \\
 
 \hline
 \textbf{\textit{Number of non-overlapping CCEs}} & 56 &56 &48&32 \\
  \hline
\end{tabular}
\label{tab1}
\vspace{-3mm}
\end{center}

\end{table}

However, the gNB transmits data on mini-slot level instead of slot level to achieve URLLC latency requirement. The UEs need to monitor PDCCH in CORSET each 2, 4 or 7 symbols instead of each slot. Thereby, the limitations in PDCCH candidates (the number of blind decodes) and CCEs reduce scheduling flexibility of the gNB for PDCCH configuration. 

In order to solve this problem, 3GPP made an agreement to increase PDCCH monitoring capability on at least the maximum number of non-overlapped CCEs per monitoring span for a set of applicable SCSs. 3GPP supports (2, 2), (4, 3) and (7, 3) as the monitoring span for SCS of 15 kHz and 30 kHz where the first number is the number of symbols between the beginning of two consecutive monitoring occasions, the second number is the number of symbols of a monitoring occasion \cite{ad15}.

The UE can be configured by the gNB to monitor PDCCH for the maximum number of PDCCH candidates and non-overlapping CCEs defined per slot as in Release 15 or for the maximum number of PDCCH candidates and non-overlapping CCEs defined per span as in Release 16. The maximum number of non-overlapping CCEs per monitoring span is the same across different spans in a slot and PDDCH might be dropped in a span in case of overbooking. Only the maximum number of non-overlapping CCEs per monitoring span for combination (7, 3) for SCS of 15 kHz and 30 kHz is defined with a value of 56 in Release 16 \cite{ad24}.  

\begin{table*}[htbp]
\caption{Scenarios for intra-UE UCI/data multiplexing}
\begin{center}
\begin{tabular}{|p{7.5em}|p{7.5em}|p{7.5em}|p{8em}|p{7.5em}|}
 \hline
  &  \textbf{\textit{URLLC SR}} &\textbf{\textit{URLLC HARQ-ACK}}&\textbf{\textit{Persistent/Semi-persistent (P/SP)-channel state information (CSI) on PUCCH}}&\textbf{\textit{URLLC PUSCH}}  \\ 
 \hline
 \textbf{\textit{URLLC SR}} &  & & &  \\
 
 \hline
 \textbf{\textit{URLLC HARQ-ACK}} & Scenario-01: reuse the Rel-15 & & & \\
   \hline
 \textbf{\textit{P/SP-CSI on PUCCH}} & Scenario-02: drop P/SP-CSI &Scenario-03: drop P/SP-CSI & & \\  
 \hline
 \textbf{\textit{URLLC PUSCH}} & Scenario-04: to be discussed &Scenario-05: reuse the Rel-15&Scenario-06: drop P/SP-CSI& \\ 
\hline
\textbf{\textit{eMBB SR}} & Scenario-07: drop lower priority SR &Scenario-08: drop SR&Scenario-09: reuse the Rel-15&Scenario-10: drop SR \\ 
\hline  
\textbf{\textit{eMBB HARQ-ACK}}  & Scenario-11: drop HARQ-ACK &Scenario-12: drop lower priority HARQ-ACK&Scenario-13: reuse the Rel-15&Scenario-14: drop HARQ-ACK \\ 
\hline   
\textbf{\textit{eMBB PUSCH}} & Scenario-15: drop PUSCH &Scenario-16: drop PUSCH&Scenario-17: reuse the Rel-15&Scenario-18: drop lower priority PUSCH \\ 

\hline   
  
\end{tabular}
\label{tab7}
\vspace{-3mm}
\end{center}

\end{table*}

\subsubsection{New DCI format}\label{IIA2}

3GPP supports a reduction of the number of bits for DCI format size compared to the size of DCI formats 0\_0 and 1\_0. The first reason of using a compact DCI is to increase reliability of DCI. A DCI with a smaller payload achieves higher reliability than a normal DCI with the same AL value. In other words, for the same reliability, a compact DCI consumes less resources because a lower AL is applied so the probability that PDCCH for a transmission cannot be transmitted in the nearest CORESET after the arrival of data due to a shortage of resources in CORESET decreases.

The presence of a new DCI format as the compact format increases the number of blind decodes that the UEs need to carry out. The gNB can configure the URLLC UEs to monitor only the compact DCI format instead of DCI format 0\_0/1\_0 and 0\_1/1\_1 so the UEs are not suffering from a growth of blind decodes. 

The target of compact DCI design agreed by 3GPP is to reduce from 10 to 16 bits \cite{ad4} by setting the number of bits in some fields to be configurable as well as reducing the sizes of some fields. The sizes of fields have been standardized by 3GPP for new DCI format 1\_2 to schedule DL transmission and new DCI format 0\_2 to schedule UL transmission\cite{ad20}, \cite{ad24}. 

New RRC parameters are configured so several fields of DCI format 0\_2 and 1\_2 are configurable without losing scheduling flexibility. Even some fields are possible to be removed (0 bit configured) in certain scenarios to reduce DCI length. For example, in DCI format 1\_2, redundancy version (RV) field is configurable from 0 bit to 2 bits compared to a fixed 2 bits in DCI format 1\_1. Similarly, hybrid automatic repeat request (HARQ) process field is configurable from 0 bit to 4 bits. Sounding reference signal (SRS) request field is configurable from 0 bit to 2 bits. Priority indicator field with 0 or 1 bit is a new field added to indicate the priority of a PDSCH scheduled. In DCI format 0\_2, open loop power control (OLPC) set indication field with from 0 to 2 bits, priority indicator field with 0 or 1 bit, invalid symbol pattern indicator field with 0 or 1 bit are new fields added to be compatible with new standards of PUSCH transmission.

\subsection{Uplink control information (UCI) enhancements}\label{IIB}

\subsubsection{Multiple PUCCHs for HARQ-acknowledgement (ACK) within a slot}\label{IIB1} 

In DL, when the gNB transmits the packets on mini-slot level, the UE is also expected to transmit feedback on mini-slot level because a fast NACK feedback on mini-slot level reduces the waiting time of the gNB and guarantees a retransmission. However, in Release 15, a UE is able to transmit only one PUCCH with HARQ-ACK information in a slot. If the UE finishes decoding process of a packet after the PUCCH resource for HARQ feedback in a slot, it must wait until the next slot to transmit feedback. Moreover, if HARQ-ACK for URLLC PDSCH occurs in the same slot as HARQ-ACK for other eMBB/URLLC PDSCHs, all the HARQ-ACK information will be multiplexed together and transmitted over the PUCCH resource indicated in the latest DL assignment. The multiplexing degrades the reliability of HARQ feedback. 

Therefore, multiple PUCCHs for HARQ-ACK within a slot are supported in Release 16 with a sub-slot-based HARQ-ACK feedback procedure where a UL slot consists of a number of sub-slots \cite{ad9}. Any sub-slot PUCCH resource is not across sub-slot boundaries and no more than one transmitted PUCCH carrying HARQ-ACK starts in a sub-slot \cite{ad20}. The number and length of sub-slots in a slot are UE-specifically semi-statically configured with two supported sub-slot configurations of 2 symbols and 7 symbols. All sub-slots in a slot have the same configuration. PDSCH-to-HARQ-feedback timing indicator is defined in unit of sub-slot to indicate the number of sub-slots from the sub-slot containing the end of the PDSCH to the sub-slot containing the start of PUCCH for HARQ feedback. The starting symbol of a PUCCH resource is defined with respect to the first symbol of sub-slot \cite{ad10}. 

3GPP agrees for Release 16 that at least two HARQ-ACK codebooks can be simultaneously constructed, intended for supporting different service types for a UE \cite{ad4}. All parameters in PUCCH configuration related to HARQ-ACK feedback can be separately configured for different HARQ-ACK codebooks. A HARQ-ACK codebook can be identified based on some PHY indications/properties  such as DCI format, RNTI, explicit indication in DCI or CORESET/search space \cite{ad7}.

\subsubsection{UCI intra-multiplexing}\label{IIB2}


In UL transmission, the number of PUCCHs transmitted by a UE in a slot is limited to 2 in Release 15. Therefore, when the UE has multiple overlapping PUCCHs in a slot or overlapping PUCCHs and PUSCHs in a slot, the UE multiplexes different UCI types (SR, HARQ-ACK) in one PUCCH as defined in Section 9.2.5 of \cite{ad22}. However, in URLLC transmission, low latency requires urgent schedules that cause an overlap of URLLC UCI with PUCCH or PUSCH of a different type services where the multiplexing between the URLLC and eMBB transmissions causes a degradation of the URLLC transmission with stringent requirements. For these cases, the behavior of the UEs must be specified to guarantee URLLC service. There are 18 scenarios listed in Table~\ref{tab7} for discussion of Release 16. The scenarios are related to the overlapping of high priority services (URLLC) with other services with the same priority (URLLC) or lower priority (eMBB). The behaviors of the UE in 17 out of 18 scenarios, except Scenario-04, are standardized in \cite{ad15} and \cite{ad20} where low priority transmission is dropped instead of being multiplexed in some scenarios. No new mechanism for Scenario-04 is agreed in Release 16. This scenario might be revisited in Release 17 if there is new update from the study of higher layer.

In the intra-collision between a high priority transmission and a low priority transmission, the non-overlapping canceled symbols are not be used for other schedulings by the UE.

In case the UE encounters the intra-collision of more than two UL transmissions, the UE resolves collision between UL transmissions with same priority then resolves collision between UL transmission with different priorities \cite{ad24}.

\subsection{PUSCH enhancements}\label{IIC}

\begin{figure}[htbp]
\centerline{\includegraphics[scale=0.17]{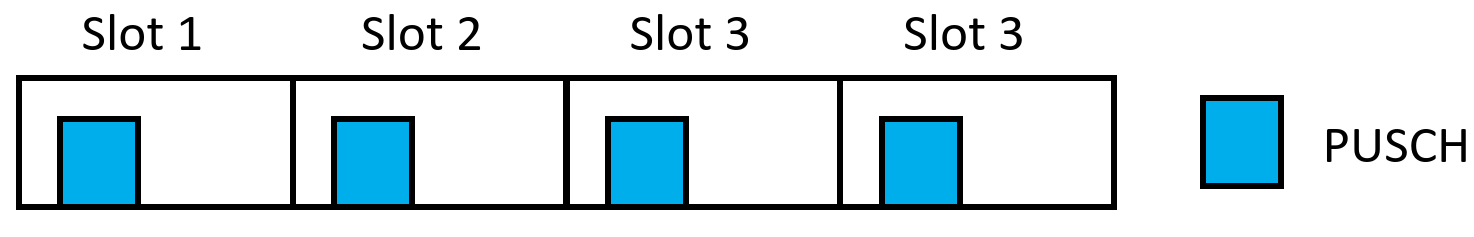}}
\caption{PUSCH repetition Type A.}
\label{fig3}
\vspace{-3mm}
\end{figure}

In Release 15, one PUSCH transmission instance is not allowed to cross the slot boundary for both DG and CG PUSCH \cite{ad13}. Therefore, to avoid transmitting a long PUSCH across slot boundary, the UE can transmit small PUSCHs in several repetitions scheduled by an UL grant or RRC in the consecutive available slots. This method is called PUSCH repetition Type A. It is also used to reduce latency of PUSCH repetitions in CG resources where a UE can be configured to transmit a number of repetitions across consecutive slots without feedback. Each slot contains only one repetition and the time domain for the repetitions of a TB is the same in those slots as shown in Fig.~\ref{fig3}. 

\begin{figure*}[htbp]
\centerline{\includegraphics[scale=0.4]{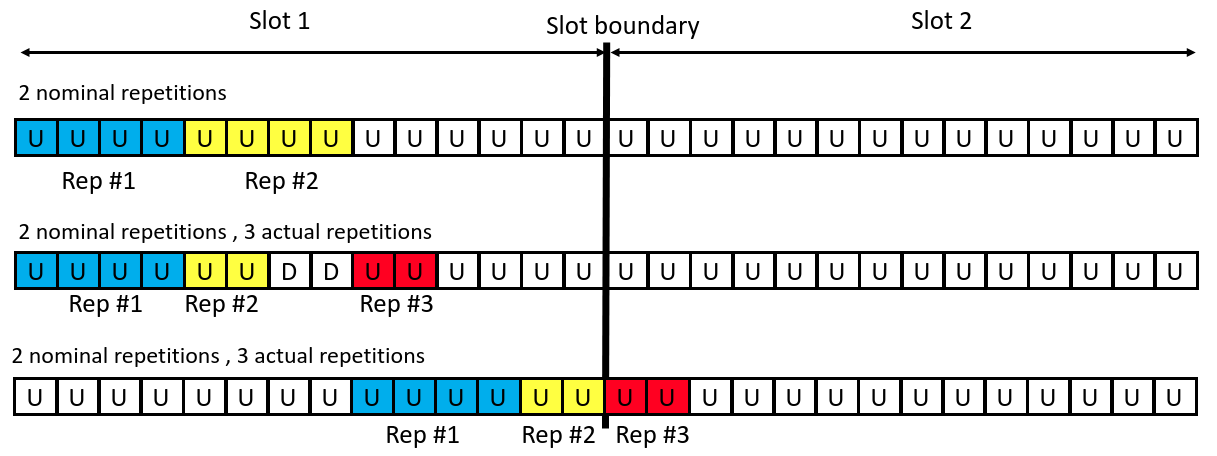}}
\caption{PUSCH repetition Type B.}
\label{fig4}
\vspace{-3mm}
\end{figure*}

However, PUSCH repetition Type A causes large time gap among the repetitions and makes the system unable to achieve URLLC latency requirement. New method called PUSCH repetition Type B in Fig.~\ref{fig4} eliminates time gap among repetitions and ensures the configured number of repetitions in the time constraint because the repetitions are carried out in the consecutive mini-slots so one slot might contain more than one repetition of a TB \cite{ad10}.

For PUSCH repetition Type B, time domain resource assignment (TDRA) field in DCI or the TDRA parameters in the type 1 configured grant indicates the resource for the first ``nominal'' repetition. The time domain resources for the remaining repetitions are derived based at least on the resources for the first repetition and UL/DL direction of symbols. The number of the repetitions represents the ``nominal'' number of repetitions. The dynamic indication of the number of repetitions for dynamic grant is jointly coded with start and length indicator (SLIV) (SLIV indicates the start symbol and length of PUSCH) in TDRA table using DCI formats 0\_1 and 0\_2 by adding an additional column for the number of repetitions in the TDRA table. The maximum TDRA table size is increased to 64 \cite{ad20}. For CG PUSCH transmission, if the number of repetitions is not included in the TDRA table, it is provided by RRC parameter repK \cite{ad24}.

If a ``nominal'' repetition goes across the slot boundary or DL/UL switching point as in Fig.~\ref{fig4}, this ``nominal'' repetition is split at the slot boundary or the switching point into multiple PUSCH repetitions. Therefore, the actual number of repetitions can be larger than the nominal number. 

For DG PUSCH with PUSCH repetition type B, if dynamic slot format indicator (SFI) is configured, two new RRC parameters are configured. The first one is to indicate a pattern of the invalid symbols for PUSCH transmission applicable to both DCI format 0\_1 and 0\_2. The second one for each DCI format 0\_1 and 0\_2 is to inform a presence of an additional bit in DCI to indicate whether the pattern configured is used or not. If the first RRC parameter is not configured, semi-static flexible symbol is used for PUSCH and segmentation only occurs around semi-static DL symbols. If the first RRC parameter is configured and the additional bit exists with value of 0, semi-static flexible symbol is used for PUSCH and segmentation only occurs around semi-static DL symbols. If the value is 1, segmentation occurs around semi-static DL symbols and invalid symbols in the pattern. If the additional bit does not exist in a DCI, segmentation occurs around semi-static DL symbols and invalid symbols in the pattern. For CG PUSCH with PUSCH repetition type B, if dynamic SFI is configured, segmentation occurs at least around semi-static DL symbols. If SFI is received for the entire duration of an actual repetition, that repetition is not transmitted if there is UL/DL conflict. If SFI is not received for at least one symbol of an actual repetition, that repetition is not transmitted if there is a conflict with a semi-static flexible symbol \cite{ad24}.

No DMRS sharing is applied among the repetitions. The maximum transport block size (TBS) does not increase compared to Release 15.

The number of actual hopping locations are 2 determined in frequency domain by reusing Rel-15 RRC parameters and equations for format 0\_1 and introducing new RRC parameters for new DCI UL format. For PUSCH repetition type B, inter-PUSCH-repetition and inter-slot frequency hoppings are supported \cite{ad24}.

For DG PUSCH with PUSCH repetition type B, the RV for the first repetition is provided by DCI and then RV is cycled across the actual repetitions following the sequence of 0, 2, 3, 1. For CG PUSCH with PUSCH repetition type B, RV is cycled across the actual repetitions following the sequence in RRC parameter repK-RV where the first repetition uses the first value in the sequence \cite{ad24}.

\subsection{Enhanced inter-UE multiplexing in UL transmission}\label{IID}

\begin{figure*}[htbp]
\centerline{\includegraphics[scale=0.5]{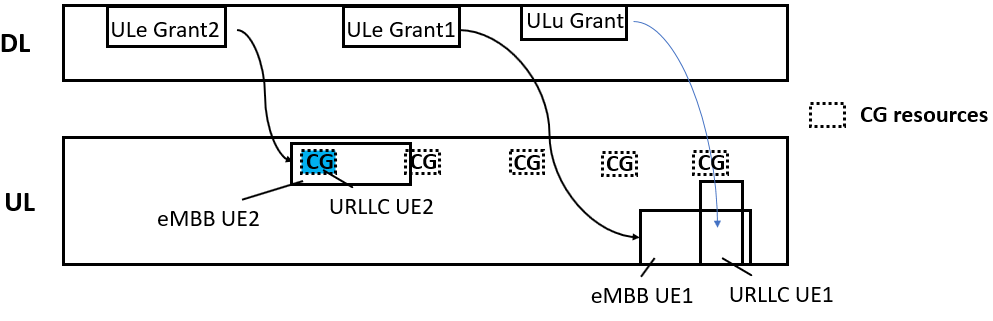}}
\caption{Two scenarios of collisions between the eMBB UEs and URLLC UEs.}
\label{fig16}
\vspace{-3mm}
\end{figure*}

To increase spectrum efficiency, latency critical communication service type and non-latency critical communication service type transmission are multiplexed in UL transmission so the gNB needs a mechanism to handle the collision and multiplexing of UL transmissions with different priorities.

The eMBB packet (low priority) collides with the URLLC packet (high priority) in UL transmission in two scenarios. In the first scenario in Fig.~\ref{fig16}, the gNB schedules UL resources to an eMBB UE1 to transmit data. After that, another URLLC UE1 also sends a SR to ask for UL resources. Due to stringent latency requirement of URLLC transmission, if no resources are available in the latency budget, the gNB must schedule the resources of eMBB transmission to the URLLC UE1 that causes a collision between the transmission of two UEs. 

In the second scenario, the gNB allocates the CG resources to the URLLC UEs. However, it does not know that the URLLC UEs have data to transmit in those CG resources or not. Thereby, the gNB also dynamically schedules the eMBB UEs to transmit in those resources in order to achieve a better resource utilization. Nevertheless, the gNB faces with a new problem of the URLLC UE2 and eMBB UE2 transmission's collision as shown in Fig.~\ref{fig16}. The degradation of URLLC transmission performance is shown by the simulations in \cite{ad7}.

Due to lack of time for the discussions in the meetings, 3GPP only supports UL cancellation indication (CI) and enhanced UL power control to handle the multiplexing between DG eMBB and DG URLLC (the first scenario) and do not support the techniques to handle the multiplexing between DG eMBB and CG URLLC (the second scenario) \cite{ad7}.

\begin{figure*}[htbp]
\centerline{\includegraphics[scale=0.55]{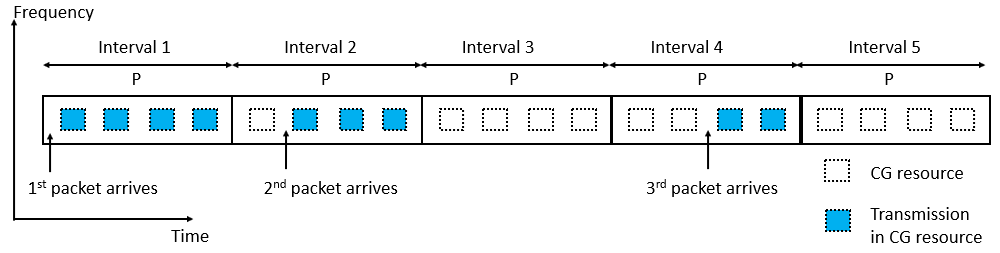}}
\caption{Less than K repetitions in CG UL transmission.}
\label{fig6}
\vspace{-3mm}
\end{figure*}

\subsubsection{UL cancellation indication}\label{IID2}

When the gNB allocates resources scheduled for the eMBB transmissions to an URLLC UE because of a strict latency requirement, it also transmits an UL CI to the eMBB UEs to ask them to stop their transmissions.

Group common DCI is supported to be used for UL CI \cite{ad10}. A new Radio Network Temporary Identifier (RNTI) is used for UL CI \cite{ad20}. The aggregation level, the number of PDCCH candidates and payload size of UL CI are configured by RRC.

It is agreed by 3GPP \cite{ad20} that PUCCH, random access channel (RACH) related UL transmission including MSG 1/3 in case of 4-step RACH, Message A (MSG A) in case of 2-step RACH cannot be cancelled by UL CI. SRS, PUSCH (DG, CG, semi-persistent (SP)) can be cancelled by UL CI. In case of PUSCH repetitions, UL CI is applied individually to each repetition overlapping the resource indicated by UL CI.

The reference time region with a duration configured by RRC starts from the given number of symbols after the ending symbol of the PDCCH CORSET carrying the UL CI where the given number of symbols is at least equal to the minimum processing time for UL cancellation. Each UL CI per serving cell has a RRC configurable field of maximum 126 bits. The reference frequency region where a detected UL CI is applicable is configured by RRC parameter \textit{frequencyRegionforCI}. The time and frequency resource for cancellation is jointly indicated by a 2D-bitmap similar to DL PI over the time and frequency partitions within the reference region \cite{ad20}. In 2D-bitmap, time domain of the overlapping regions is divided into 1, 2, 4, 7, 14 or 28 partitions (configured by RRC parameter \textit{timeGranularityforCI}) mapping to the corresponding number of bits. In TDD configuration, the DL symbols indicated by RRC parameter \textit{tdd-UL-DL-ConfigurationCommon} are excluded when the partitions of reference time region are chosen. The number of partitions in frequency domain of the overlapping regions is the division of the total number of indication bits and the number of bits indicating time domain \cite{ad24}.

The monitoring periodicity of the UE for UL CI is configurable to maximum 5 slots by the gNB. The UE can monitor UL CI in more than one occasions in a slot with periodicity of 2, 4 and 7 symbols configured by RRC parameter \textit{timedurationforCI} \cite{ad24}. This means that mini-slot-based monitoring might be used instead of slot-based monitoring \cite{ad4}.

To reduce the UE monitoring for UL CI and satisfy UE capability, the number of ALs and/or candidates for UL CI monitoring should be limited. The UE processing time requirement for UL CI is defined in Rel-15 \cite{ad20}.

When the UE detects UL CI from the gNB, it stops the transmission without resuming. The scheme that allows the UE to resume the unfinished transmission after the region indicated by UL CI is not supported. There is an exception when SRS can still be transmitted on the non-cancelled symbols \cite{ad15}.

\subsubsection{Enhanced UL power control}\label{IID3}

Besides using UL CI in eMBB and URLLC multiplexing, the gNB has a second option by using power control scheme. The gNB sends a signal to increase power level of URLLC PUSCH transmission in the overlapping region with the eMBB transmission. It helps the UE operate in a higher SNR and compensates the effect from the interference of the eMBB transmission. However, power boosting is not applicable to the power limited UEs.

For DG PUSCH, open-loop parameter sets based on scheduling DCI (UL grant) without using scheduling request indicator (SRI) field is supported to control transmission power \cite{ad15}. One or two bits in UL grant are used to indicate the open loop power control parameter set when one new RRC parameter containing one additional \textit{P0-PUSCH-Set} per SRI is configured with one or two P0 values. The UE uses P0 value in \textit{P0-PUSCH-AlphaSet} as Release 15 if the indication is ``0'' in case of 1-bit indication or ``00'' in case of 2-bit indication or if the indication field is not present in UL grant. The UEs uses the first P0 value in \textit{P0-PUSCH-Set} if the indication is ``1'' in case of 1-bit indication or ``01'' in case of 2-bit indication. The UEs uses the second P0 value in \textit{P0-PUSCH-Set} if the indication is ``10'' \cite{ad20}, \cite{ad24}.

\subsection{Enhanced UL configured-grant transmission}\label{IIE}

In Release 15, the UE is able to transmit blindly repetitions without waiting for feedback from the gNB. However, the UE is only allowed to transmit the repetitions in one HARQ process interval to avoid the confusion between the initial transmission and the retransmissions at the gNB. If the gNB misses the first transmission and only detects the retransmissions in a different HARQ process to that of the first transmission, the gNB will use the wrong UE HARQ identity in the UL grant to schedule a retransmission. Due to this constraint, the UE must stop to carry out the repetitions if it reaches the boundary of a HARQ process even if it still has not transmitted all repetitions configured as the second and the third packet in Fig.~\ref{fig6}. In consequence, reliability of URLLC transmission reduces.

In \cite{ad13}, it is agreed that multiple active CG configurations for a given Bandwidth part (BWP) of a serving cell should be supported at least for different services/traffic types and/or for enhancing reliability and reducing latency. Ensuring the configured number of repetitions is included in the target of enhancing reliability and reducing latency of multiple active CG configurations.

The UE chooses the configuration with the earliest starting point to transmit data. It ensures that data is always transmitted at the beginning of a HARQ process interval and the configured number of repetitions is transmitted before reaching the HARQ process boundary.

The maximum number of UL CG configurations per BWP of a serving cell is 12 as supported by 3GPP in \cite{ad10}. One UE might have multiple configurations and one configuration might be shared among several UEs. The gNB send RRC (for both type 1 and type 2 CG transmission) or DCI (only for type 2 CG transmission) to activate or release the configurations. It is agreed in \cite{ad9} that separate RRC parameters for different CG configurations (for both type 1 and type 2 CGs) for a given BWP of a serving cell are supported. 

3GPP also agreed to support separate activation by DCIs (DCI format 0\_0, 0\_1 and 0\_2 \cite{ad20}) for different CG Type 2 configurations for a given BWP of a serving cell. There are maximum 4 least significant bits of HARQ process number (HPN) field to indicate which configuration is activated. Joint activation in a DCI is not supported \cite{ad15}. 

In activation, only separate activation is allowed. However, in release of the active configurations, both separate release and joint release are allowed as the agreement in \cite{ad10}. The gNB sends the Release DCI (DCI format 0\_0, 0\_1 and 0\_2) to indicate which CG configurations are released.  There are maximum 4 least significant bits of HPN field in the Release DCI to identify the configurations. One state can be used to indicate one configuration or multiple configurations released. In case of no higher layer configured states, separate release is used where the release corresponds to the CG configuration index indicated by the Release DCI \cite{ad15}.

\section{Conclusion}\label{III}

URLLC is specified as one of the three key services of 5G New Radio. In order to satisfy URLLC requirements, new techniques in physical layer design are demanded. This article introduced the problems in URLLC physical layer design and a set of techniques specified in Release 16 in order to solve these problem. The new features in Release 16 improve the performance of URLLC transmission and have a role as a bridge to the next evolution of URLLC. The first objective of the next 3GPP release is physical layer feedback enhancements covering UE feedback enhancements for HARQ-ACK and CSI feedback enhancements to allow more accurate MCS selection. The second objective is to guarantee Release 16 feature compatibility with unlicensed band URLLC/industrial internet of things (ITOT). Another objective is intra-UE multiplexing based on the work of Release 16.

\section*{Biographies}

Trung-Kien Le received MCS. degree specializing in Mobile Computing Systems from EURECOM, France. Currently, he is a PHD student in Sorbonne University and EURECOM. His research interest is Ultra Reliable Low Latency Communication in 5G New Radio.

Umer Salim received his Ph.D. and M.S. degrees, specializing in communication theory and signal processing from Eurecom and Supelec, France, respectively. He is currently working at TCL Communications as 5G Systems Architect. 

Florian Kaltenberger received his Dipl.-Ing. degree and his Ph.D. degree both in technical mathematics from the Vienna University of Technology in 2002 and 2007, respectively. He is an assistant professor at the Communication Systems Department of Eurecom, Sophia-Antipolis, France.

\end{document}